\documentclass[pre,amsmath,amssymb, twocolumn, showpacs]{revtex4}
\usepackage{dcolumn}
\usepackage{bm}
\usepackage{graphicx}
\usepackage{hyperref}
\usepackage{mathptmx}
\usepackage{subfigure}
\usepackage[utf8]{inputenc}
\usepackage{bbold}
\usepackage{color}

\newcommand{\RR}{\mathbb{R}}
\newcommand{\ZZ}{\mathbb{Z}}

\newcommand{\Eperp}{E_\bot}

\makeatletter

\begin{document}

\title{Horizons and free path distributions in quasiperiodic Lorentz gases}
\author{Atahualpa S.~Kraemer}
\email{akraemer@thphy.uni-duesseldorf.de}
\affiliation{Heinrich-Heine Universit\"at D\"usseldorf,
Universit\"atsstrasse 1,
D-40225 D\"usseldorf, Germany
}
\author{Michael Schmiedeberg}
\affiliation{Heinrich-Heine Universit\"at D\"usseldorf,
Universit\"atsstrasse 1,
D-40225 D\"usseldorf, Germany
}
\author{David P.~Sanders}
\affiliation{Departamento de F\'isica, Facultad de Ciencias,
Universidad Nacional Aut\'onoma de M\'exico,
Ciudad Universitaria, M\'exico D.F. 04510, Mexico
}
 
\date{\today}
\begin{abstract}
We study the structure of quasiperiodic Lorentz gases, i.e., particles bouncing elastically off fixed obstacles arranged in quasiperiodic lattices. By employing a construction to embed such structures into a higher-dimensional periodic hyperlattice, we give a simple and efficient algorithm for numerical simulation of the dynamics of these systems.
This same construction shows that quasiperiodic Lorentz gases generically exhibit a regime with infinite horizon, that is, empty channels through which the particles move without colliding, when the obstacles are small enough; in this case, the distribution of free paths is asymptotically a power law with exponent -3, as expected from infinite-horizon periodic Lorentz gases.
For the critical radius at which these channels disappear, however, a new regime with locally-finite horizon arises, where this distribution has an unexpected exponent of -5, 
 previously observed only in a Lorentz gas formed by superposing three incommensurable periodic lattices in the Boltzmann-Grad limit where the radius of the obstacles tends to zero.
\end{abstract}


\maketitle

\section{Introduction}

The Lorentz gas (LG) model was proposed by 
Lorentz \cite{lorentz1905motion} as a model of a completely ionized gas to study the conductivity of metals.
During the last decades, Lorentz gases have became popular among mathematicians, as key models in probability theory and dynamical systems~\cite{dettmann2014diffusion}. At the same time, numerous works in physics use modified versions of the Lorentz gas to study dynamical and statistical properties of systems
with periodic~\cite{77, 80, 87, 88,90, 76, 78, 144, 100, 8, 56} and random~\cite{82,103, 102, 81,113, 130} distributions of obstacles in one~\cite{81, wennberg2012free}, two~\cite{132}, three~\cite{130, gilbert2011diffusive}, and higher dimensions~\cite{100, 8, dettmann2014diffusion, gilbert2011diffusive}. 
It has been shown heuristically, numerically~\cite{8, 100}, and rigorously~\cite{nandori2014tail, dettmann2014diffusion, marklof2014superdiffusion, marklof2015invariance} that $m$-dimensional Lorentz gases generically present weak super diffusion if there are channels of dimension $m-1$ (also called principal horizons~\cite{dettmann2014diffusion}) in which particles can move freely for infinite time i.e., the mean squared displacement $\langle \Delta x(t)^2 \rangle = \langle x(t)^2 \rangle - \langle{x(t)}\rangle^2  \sim  t \log(t)$, where $\langle X \rangle$ is the ensemble average of $X$. This situation is generic for periodic Lorentz gases~\cite{dettmann2014diffusion} and it has been suggested for quasiperiodic Lorentz gases~\cite{PRL-Ata-David}; however, the proof of a generic situation in quasiperiodic systems is still missing. 

On the other hand, in solid state physics the exploration of aperiodic structures, such as quasicrystals, has become increasingly important, since these systems exhibit a number of surprising effects, such as phasons \cite{socolar1986phonons, PhysRevLett.108.218301}. 
Quasicrystals are structures with long-range order, but no
translational symmetry \cite{93}, first found experimentally by Shechtman in a metalic alloy with a diffraction pattern with 10-fold symmetry~\cite{67}.
Since their discovery, quasicrystals have been produced with many different materials \cite{4, 6, 13, 15, 109, tsai2000alloys, PhysRevB.62.R14605}. 
Quasiperiodic arrays have also been found in other contexts, e.g.,
liquid quasicrystals~\cite{143}, auto-assemblies of
nanoparticles~\cite{assemble-binary-nanoparticlq}, virus colonies~\cite{PhysRevLettVirus}, and photonic quasicrystals \cite{zoorob2000complete, 146}.
Furthermore, quasicrystals have been observed in nature~\cite{140,141}. In simulations,
quasicrystalline structures have been found as cluster quasicrystals \cite{Barkan2014} or in hard tetrahedral systems~\cite{74}, where a first-order phase transition was observed, as confirmed in experiments \cite{154}.

In spite of the relevance of quasiperiodic systems, quasiperiodic Lorentz gases have only recently been investigated~\cite{wennberg2012free, PRL-Ata-David, marklof2014free, 16, marklof2014visibility}, having previously been 
proposed as an open problem in the theory of dispersing billiards \cite{16}. 

In particular, the distribution of free paths has been studied in the Boltzmann-Grad limit~\cite{wennberg2012free, marklof2014free}, where the radius of the obstacles tends to zero. In this limit, it has been proved that the distribution should be similar to the periodic case \cite{marklof2014free}, i.e., the probability density of free paths of length $\ell$ should decay asymptotically as a power law $\ell^{-\alpha}$, with exponent $\alpha=3$ \cite{boca2007distribution, nandori2014tail}. 

\begin{figure}
\centering
\includegraphics*[width=245pt]{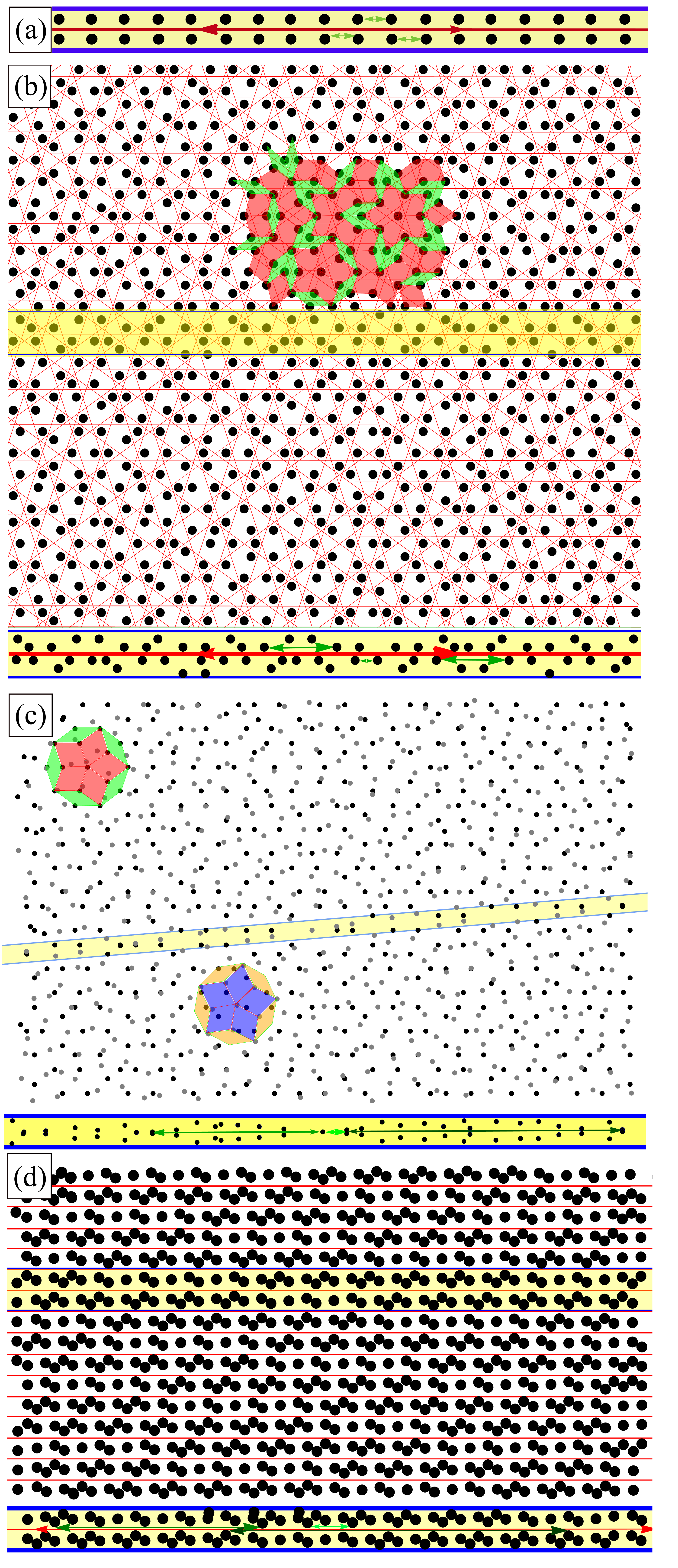}

 \caption{(Color online) Periodic and quasiperiodic arrays of scatters.  
 (a) A channel in a square lattice of obstacles. 
 (b) Different channels (thin red (strong gray) lines) in the Penrose Lorentz gas. At the bottom are shown trajectories in the direction of a horizontal channel. 
 (c) Combining two Penrose Lorentz gases, the second rotated by an angle $\frac{\pi}{20}$, blocks the channels. At the bottom some trajectories are shown in a direction between two of the channels of the two Penrose arrays. 
 (d) Quasiperiodic array that results from projecting a 3D lattice into a 2D subspace. At the bottom we show a channel and some trajectories in the direction of the channel.}
\label{fig: periodic-quasi}
\end{figure}

Consider a periodic Lorentz gas and a particle that moves outside a channel, but close to the direction of this channel. In general, the length of the free path of this particle will be bounded by the lattice spacing; see fig.~\ref{fig: periodic-quasi}(a)). When channels are blocked in a \emph{periodic} Lorentz gas, the distribution of free path lengths becomes bounded above. It is not a priori clear if the same will hold for quasiperiodic lattices; see fig. \ref{fig: periodic-quasi}(b--d). In fact, we show in this paper that quasiperiodic Lorentz gases generically have a regime with so-called \emph{locally-finite} horizon, where the width of the largest channel is zero, but there is \emph{no upper bound} on the free path length \cite{troubetzkoy2010typical}. This situation occurs only precisely at a critical radius $r = r_c$, defined such that channels are present when $r < r_c$, and absent when $r > r_c$, i.e. in the limit when the width of the widest channel tends to $0$.

As mentioned above, in an $m$-dimensional Lorentz gas, if there are $(m-1)$-dimensional channels present, it is expected that the distribution of free path lengths is a power law with exponent $\alpha=3$ \cite{bouchaud1985numerical, 8, 100}, and that the diffusion has a logarithmic correction to the mean square displacement. 
One of the most studied examples of a system with locally-finite horizon is the random Lorentz gas, in which the obstacles are often distributed with positions following a Poisson distribution (if overlapping is allowed). In this case, the distribution of the length of the free paths seems to be 
exponential, at least in the Boltzmann-Grad limit~\cite{marklof2014low}. It is natural to ask what distribution is found in quasiperiodic Lorentz gases. 

There are many other possible random distribution of obstacles, for example, when the scatterers are of finite size, they are often considered to be non-overlapping and hence not Poisson. We are not aware of any numerical investigations of the free path distribution for this case at a high density, but in Ref.~\cite{kruis2006systematic} this has been calculated for both overlapping and non-overlapping obstacles in the low density limit. In both cases the distribution of free path length has an exponential tail.

This paper is organized as follows: In Section~\ref{model}, we define a quasiperiodic Lorentz gas and we summarize the procedure to embed a quasiperiodic potential into a 
higher-dimensional periodic potential. In Section~\ref{Finite and infinite} we define finite, infinite, and 
locally-finite horizons in Lorentz gases and we prove the generic existence of channels for quasiperiodic Lorentz gases. We then show that quasiperiodic Lorentz gases have a locally-finite horizon for $r = r_c$. In Section~\ref{results}, we measure numerically the free path length distribution, obtaining, for the locally-finite regime, a power law with an unexpected exponent $\alpha=5$, which we confirm with heuristic arguments. We finish with conclusions in Section~\ref{conclusions}.

\section{Model and methods}
\label{model}

\subsection{Lorentz gas}

A Lorentz gas (LG) consists of an ensemble of non-interacting point particles moving in an array of fixed obstacles, usually spheres, placed at the vertices of a lattice in $\RR^{m}$. Each particle undergoes free motion until it collides with a scatter, and is then reflected elastically. If the lattice is quasiperiodic, then the Lorentz gas is also called quasiperiodic. There are several methods to produce quasiperiodic arrays, not all of which producing the same tiling \cite{lamb1998canonical}; one of the most popular is the projection method \cite{95,93,96}. Reversing this method, it is possible to simulate and analyze the dynamics of a quasiperiodic LG as a periodic billiard, as two of the current authors previously showed~\cite{PRL-Ata-David}. Throughout this paper, we take the interaction potential to be that of the hard sphere.

One of the main interests in studying a Lorentz gas consists of measuring its diffusivity, i.e., how fast particles disperse through the system, characterized by the variance, or mean squared displacement, $\langle \Delta x(t)^2 \rangle$, of an initial cloud of particles as a function of time, $t$, where the ensemble average is defined by averaging with respect to the uniform measure over the unit cell in the higher-dimensional periodic system. 

It has been shown, numerically~\cite{56, 8} and analytically~\cite{90}, that the periodic version of these models can exhibit weak super-diffusion:
\begin{equation} 
\langle {\Delta x^2}(t) \rangle \sim D t \log(t/\tau), 
\label{eq: superdifusion} 
\end{equation} 
where $D$, the super-diffusion coefficient,
is a constant (for a given system) that depends on the geometry of the lattice on which the obstacles are positioned and the obstacle radius, and $\tau$ is the average time that a particle stays in a cell of a given size.

This occurs in the presence of the highest possible dimension of \emph{channels} in the structure, i.e., subspaces of the system in $\mathbb{R}^m$, such that the dimension of the set of infinite directions of the channel is $m-1$, and which are devoid of obstacles. This happens generically in periodic LGs if the obstacles are small enough; however, it does not happen in random LGs. 

Simulations of the mean square displacement in quasiperiodic LGs close to the locally-finite horizon regime suggest that such systems have normal diffusion~\cite{PRL-Ata-David}. Nonetheless, the simulation data may not be sufficient; in particular, calculating logarithmic corrections numerically is subtle \cite{logcorrectionCTD}. 

As an alternative, we examine the distribution of free path lengths obtained with a locally-finite horizon. It is expected that a power law with exponent $\alpha=3$ for this distribution corresponds to weak super-diffusion (logarithmic correction), while a smaller exponent instead corresponds to normal diffusion~\cite{100}.

\subsection{Periodization of quasiperiodic potentials}
\label{periodizando}

We summarize the method for constructing finite-range quasiperiodic potentials introduced in \cite{PRL-Ata-David}. The main idea is to produce an $n$-dimensional periodic potential, with $n > m$, based on the projection method \cite{95,93,96}, with non-interacting classical particles moving inside.  The initial conditions are constrained such that the dynamics of particles in the periodic potential will reproduce the dynamics in the $m$-dimensional quasiperiodic potential: the initial velocities must lie in the $m$-dimensional physical subspace $E$ onto which points of the higher-dimensional periodic lattice are projected. $E \subset \RR^n$ is a \emph{totally irrational} subspace, i.e., such that $E \cap \ZZ^n = \{0\}$.

\begin{figure}
\centering

\noindent\begin{tabular}{@{\hspace{0.0em}}c@{\hspace{0.0em}}c@{\hspace{0.0em}}}
   \includegraphics[height=0.6\linewidth]{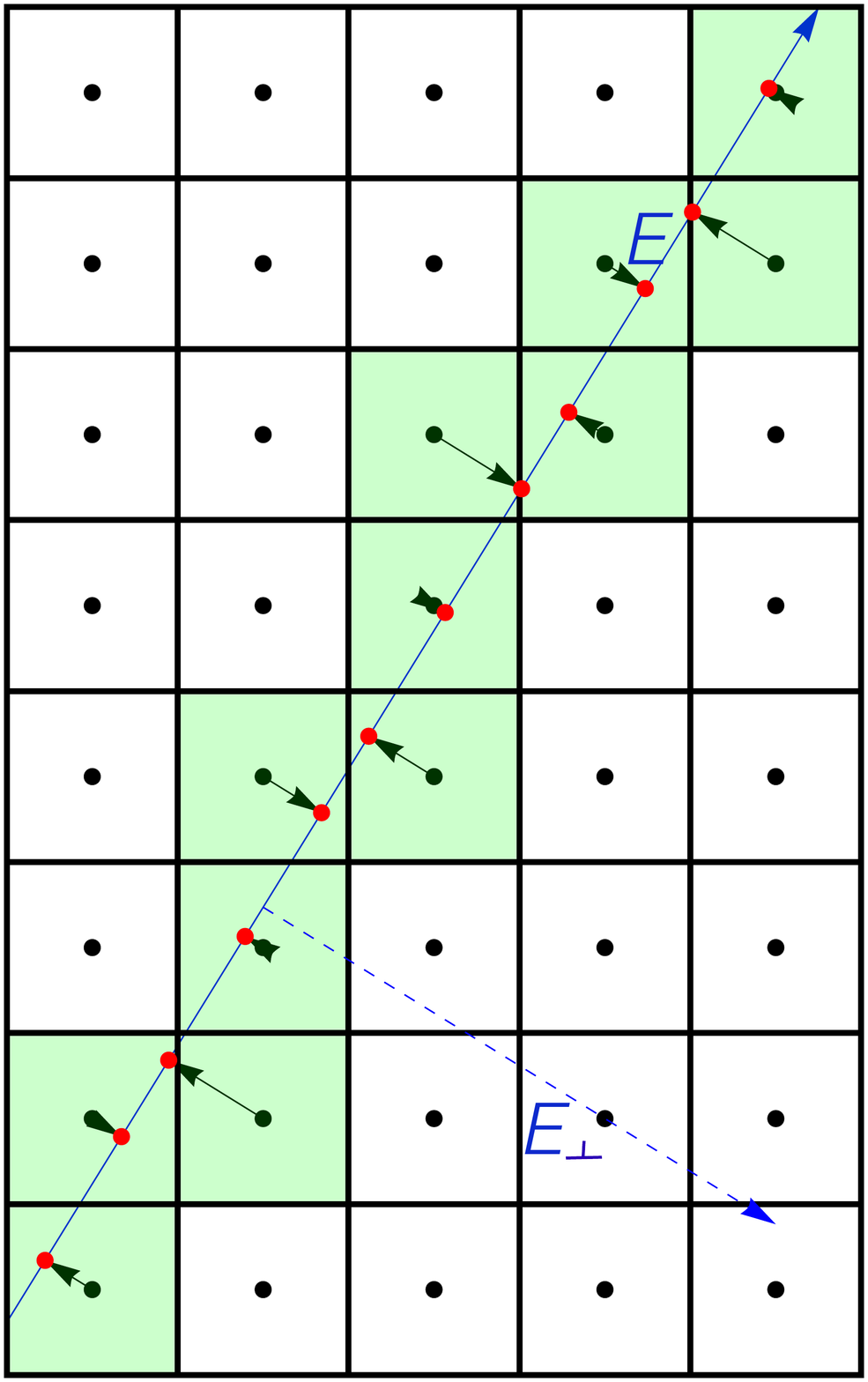} &
   \includegraphics[height=0.6\linewidth]{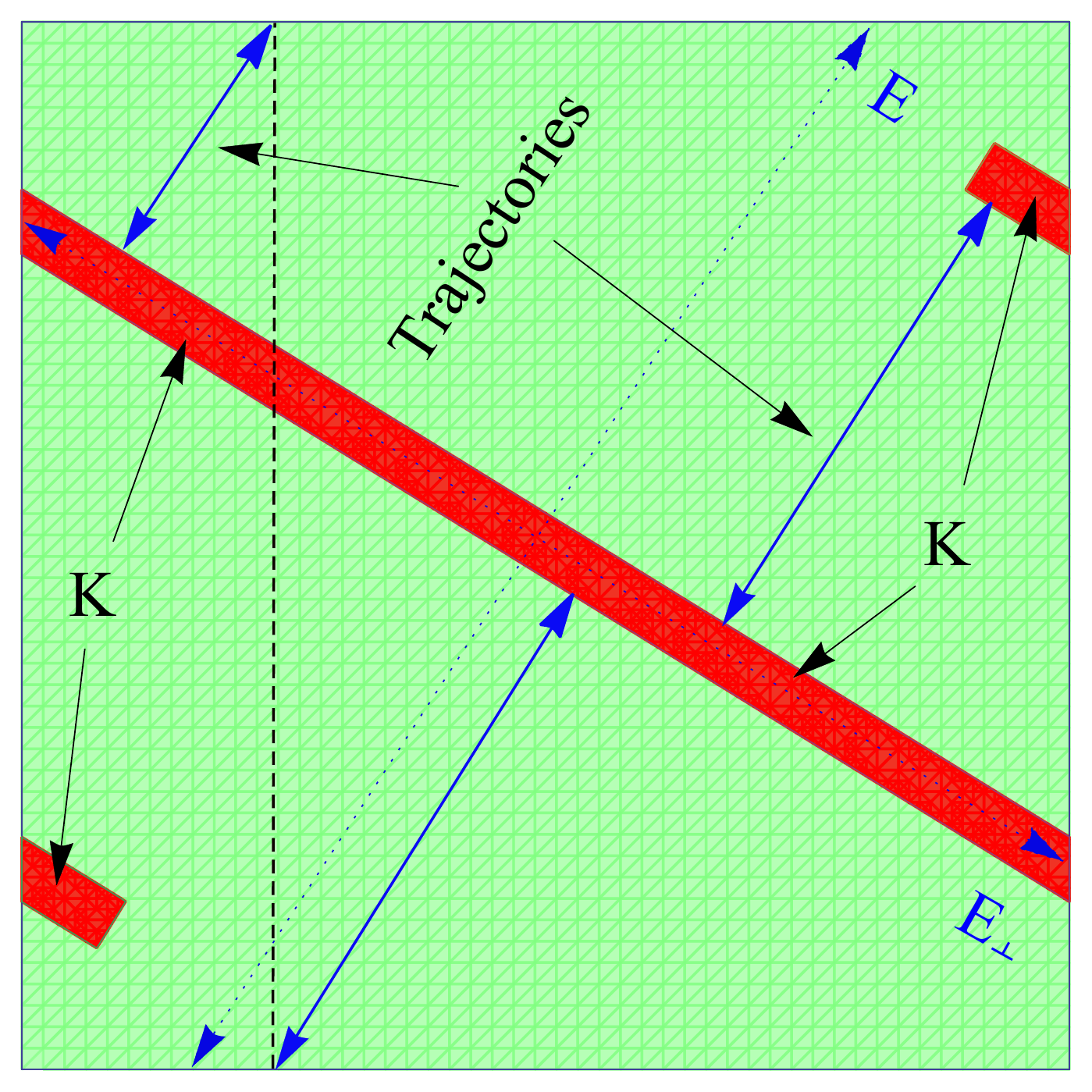} \\
   (a) & (b)
\end{tabular}
 
\caption{(Color online) (a) Projection method used to produce the Fibonacci chain. $E$ is the physical space, where the particles are projected to produce the Fibonacci chain. (b) Periodization of the Fibonacci chain; particles can move in only two directions.}
\label{fig:projection method}
\end{figure}

Fig.~\ref{fig:projection method}(a) shows, as an example of the projection method, the quasiperiodic  Fibonacci chain, constructed by projecting a 2-dimensional periodic lattice onto a 1-dimensional totally irrational line $E$, i.e., a line with irrational slope. Figure~\ref{fig:projection method}(b) shows a 2-dimensional periodic potential, which represents the Fibonacci chain if the particles are constrained to move only parallel to $E$. 

In general, we can construct periodic potentials in higher dimension that are equivalent to a two- or three-dimensional quasiperiodic array in $E$. For example, the Penrose tiling can be embedded in a 5-dimensional periodic potential or the icosahedral array can be embedded in a 6-dimensional periodic potential \cite{senechal1996quasicrystals}. To do so, we proceed as follows: 

\begin{enumerate}
\item Take a unit hypercube $C$ of dimension $n$, corresponding to a Voronoi cell of the periodic lattice.
\item Translate the space $E$ so that it passes through the center $c$ of the hypercube $C$, which we take as the origin.
\item Project $C$ onto $E_\perp$, the subspace orthogonal to $E$ (of dimension $n-m$) that also passes through $c$. Call the resulting projected object $W$. 
\item Apply periodic boundary conditions to $W$, by translating those parts of $W$ that lie outside $C$, to produce a ``periodized'' object $K$ inside $C$. 
\item Apply the $m$-dimensional potential (for example, the hard-sphere potential) in the direction of the hyperplane $E$, using $K$ as the axis of the potential. In the orthogonal direction, the potential is $0$. 
\end{enumerate}

We call this procedure to embed a quasiperiodic system into a higher-dimensional periodic one the \emph{periodization} of the system.

\section{Horizons in quasiperiodic Lorentz gases}
\label{Finite and infinite}

In this section, we define finite, infinite and locally-finite horizons, and we prove the generic existence of channels in quasiperiodic Lorentz gases, as well as the locally-finite horizon regime. 

\subsection{Generic existence of channels in quasiperiodic Lorentz gases}

The construction described in the previous section was originally designed to allow efficient numerical simulation of quasiperiodic Lorentz gases. Nonetheless, it also provides a powerful tool to analyze the geometric structure of these systems 
\cite{PRL-Ata-David}: here we use it to prove the generic existence of channels in quasiperiodic LGs; specific cases were studied in \cite{PRL-Ata-David}.

The periodized model is a periodic LG in a higher dimension, in which the obstacles are no longer spheres, but are now $n$-dimensional cylinders (together with the constraint mentioned above on the initial velocities of the particles). 
If the radius of the obstacles is small enough, then we expect that there will be channels of dimension $n-1$ in the $n$-dimensional periodic LG. 
We need only prove that these channels are not all parallel to the plane $E$;
if so, then there are channels of dimension $m-1$ in the quasiperiodic LG, since the intersection of a subspace of dimension $n-1$ with a subspace of dimension $m \leq n$ is generically a subspace of dimension $m-1$; for example, the intersection of two planes in 3D is generically a line. 

\begin{figure}
\centering
\includegraphics[width=240pt]{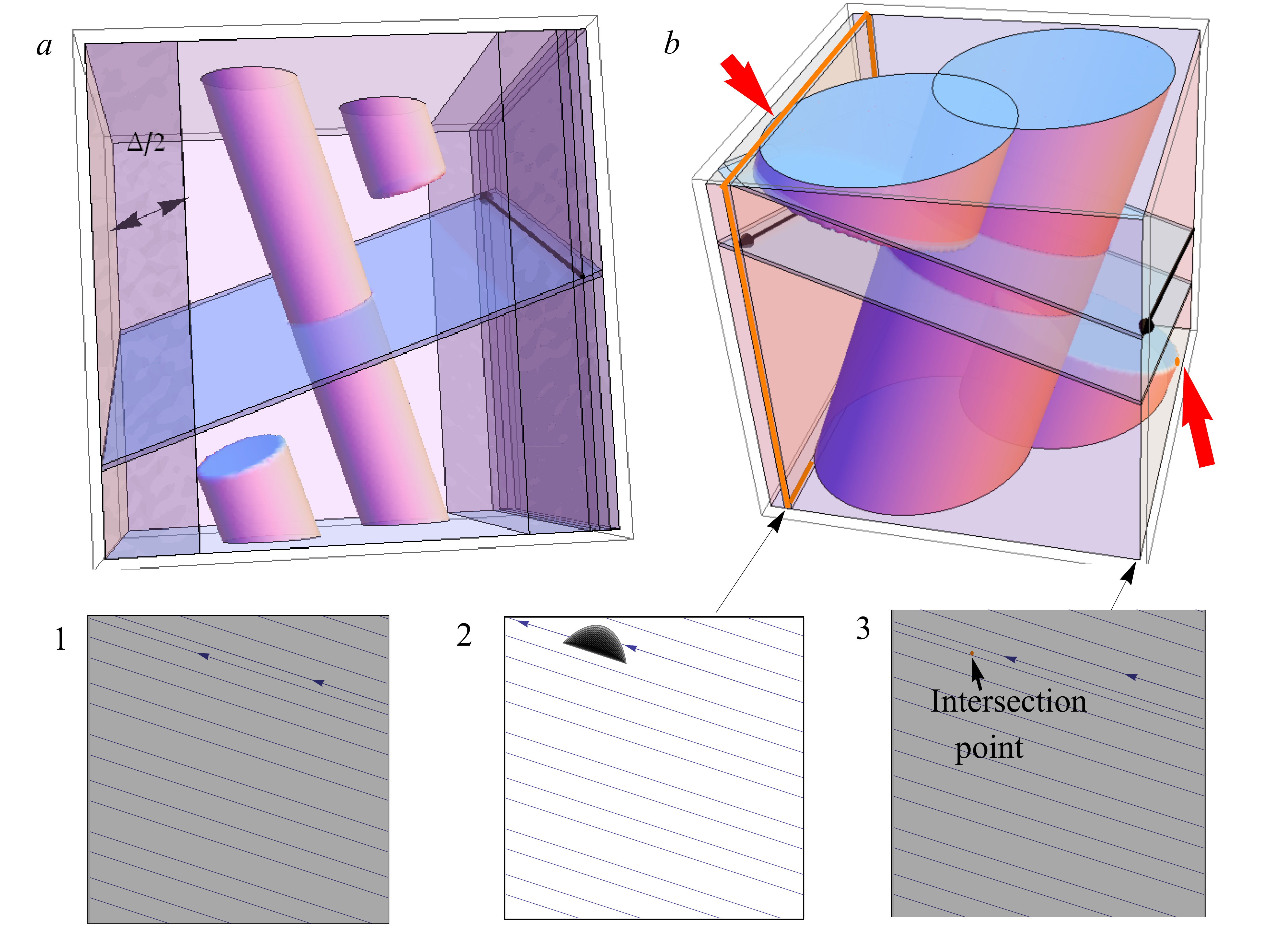} 
 \caption{(Color online) Trajectory of a particle constrained to move in plane parallel to one of the faces of the cube, in the case that the plane $E$ is totally irrational. The trajectory of the particle inside a channel represented in (a) fills densely the face of the cube, represented in (1) as a gray background with part of the trajectory highlighted. If the channels become blocked (b), the trajectory can still be dense in the face of the cube (3) or becomes finite if the particles moves in a plane parallel to the face of the cube (2), since it collides with the part of the ellipse (shown in black) that is the intersection of the cylinder with this plane. The thick (red) arrows on figure (b) show the intersection between the planes. The bottom arrow shows only a point while the other shows a larger area of intersection.}
 \label{fig:caricatura2}
\end{figure}

To show the existence of these 
$(n-1)$-dimensional channels that are not parallel to $E$, note that if a face of $C$, which is an $(n-1)$-dimensional hyperplane, does not touch the obstacle, then there will be a channel with this property, since the plane $E$ is totally irrational, so that it cannot be parallel to any face of $C$. Thus, the intersection of the plane $E$ (of dimension $m$) and this face will produce a subspace of $E$ with dimension $m-1$, without any obstacle; that is exactly the definition of channel. This happens generically since $K$ has the same dimension as the orthogonal space to $E$, namely $n-m < n$, and its length is bounded by the length of the hypercube such that the intersection between the hypercube and $W$ intersects the same number of faces as $K$; see 
Figs.~\ref{fig:caricatura2}(a) and \ref{fig:caricatura2}(1). In this case, it is not possible that $K$ intersects all the faces; indeed, there are exactly $2m$ faces that it does not intersect, $m$ of them orthogonal, giving exactly $m$ channels if the obstacle is small enough.  

Therefore, we expect weak super-diffusion if the obstacles are small enough, which agrees with numerical results founded in \cite{PRL-Ata-David}. However, the numerical results are not completely convincing, especially when the obstacles are very small. The logarithmic correction to the mean square displacement is difficult to observe numerically even in periodic systems~\cite{logcorrectionCTD}. This problem persists in the quasiperiodic case, but is even more challenging, since there is an additional effect that results in slow convergence to the logarithmic correction, as we will see in the following.

\subsection{Locally-finite horizon}

Periodic Lorentz gases can be classified into systems with finite or infinite horizon, according to whether the free path length is bounded above or can be infinite.
In contrast, random Lorentz gases are typically in the locally-finite regime, in which the probability to have unbounded free paths is $0$, but the length of free paths may be arbitrary large~\cite{troubetzkoy2010typical}. Furthermore, if we fix an arbitrary direction, the free path in that direction is still bounded, with probability $1$. Thus, we will say that a system has locally-finite horizon if the free path length is not bounded above, but for any fixed direction the probability of choosing a point with an unbounded trajectory in that direction is $0$. We will show here that quasiperiodic Lorentz gases can exhibit all three of these regimes: finite horizon, locally-finite horizon and infinite horizon, depending on the size of the obstacles. 

\begin{figure}
\centering
\includegraphics[width=245pt]{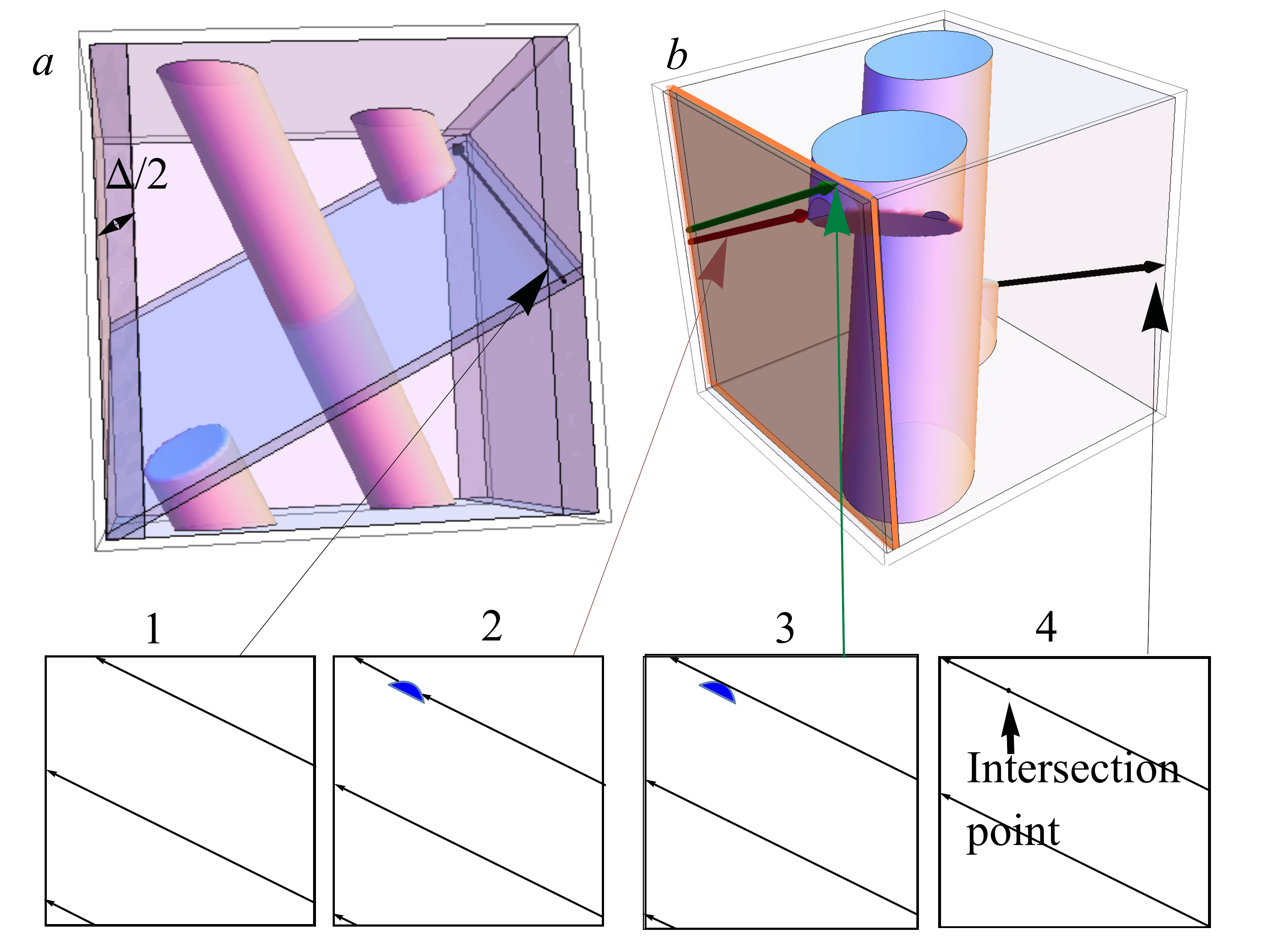} 
 \caption{(Color online) Trajectory of a particle constrained to move in a plane parallel to one of the faces of the cube, in the case that the plane $E$ is totally rational: (a) in the presence of a channel of width $\Delta$; (b) when the channels are blocked. Below are shown trajectories of particles moving in a face of the cube. (1) A particle in a channel; (2) and (3) show two cases where the particle touches tangentially the obstacle; (4) a particle that collides with an obstacle. The cases (3) and (4) are contained in the plane marked in a dark colour in figure (b), while figure (4) is contained in one of the faces of the cube. }
\label{fig: faces}
\end{figure}

Suppose that the plane $E$ of dimension $m$ in the periodization of a quasiperiodic array is totally \emph{rational}, i.e., $E \cap \ZZ^n$ is a lattice of rank $m$, rather than totally irrational. In this case, the periodized system represents a periodic Lorentz gas, rather than a quasiperiodic one. Note that a trajectory in this system will not fill densely the available volume. For example, consider a particle with initial velocity and position in the intersection of the plane $E$ and a plane $F$ that contains one of the faces of the cube that is not touched by the obstacle if the obstacle is small enough, i.e., a channel. Then, as shown in fig.~\ref{fig: faces}, the whole trajectory will be represented by a finite number of segments. If we increase the radius of the obstacle, at some moment the obstacle will intersect this face. The intersection of the obstacle and the plane $F$ will produce an $(n-1)$-dimensional object. If we continue growing the obstacle, this $(n-1)$-dimensional  object  will continue 
growing until it intersects the trajectory. At this point, the channel becomes blocked, and the free path  becomes finite and bounded. That is, the horizon changes from an infinite to a finite horizon, without passing through a locally-finite horizon. 

The situation is similar if, instead of growing the obstacle, we continuously displace the plane $F$ to the plane $F_{\lambda}$, keeping it parallel to the face of the cube,  where 
$\lambda$ denotes the distance between the plane and the face of the cube. We can identify the region where particles with velocity $v$ in the direction of the intersection of the planes $F_{\lambda}$ and the plane $E$ can move freely by seeing where the intersection of the plane $F_{\lambda}$ and the obstacle does not intersect the trajectory of the particles constrained to move in this plane. This region is the channel with direction $v$. 

Now, consider the quasiperiodic Lorentz gas, i.e., when the plane $E$ is totally irrational. Following the same procedure, we produce first a trajectory that fills densely the face of the cube. If the intersection between the plane $F$ and the obstacle is non-empty, the trajectory will be finite; see Fig.~\ref{fig:caricatura2}. This time, however, the length of the trajectory depends on the size of the $(n-1)$-dimensional object produced by this intersection. Consider the limiting case in which the obstacle is of the critical size such that it exactly touches the plane $F$, so that the intersection  between the plane $F$ and the obstacle consists of exactly one point on the plane $F$. Then the channel associated to this face has measure $0$, and the length of the free paths constrained to move in the planes $F_\lambda$ depend on $\lambda$, with no upper bound. Thus, in the case when all channels become blocked and at least one of them is in this limiting case, the system has a locally-finite horizon.

This can explain why it is more difficult to see numerically the behavior of the mean square displacement of the form $D t \log(t/\tau)$ in quasiperiodic Lorentz gases than in the periodic case. The coefficient $D$ depends on the width of the channel \cite{100}, but this width is, in some sense, a decreasing function of time $t$ for quasiperiodic systems: for any time $t$, there is a volume of the billiard (width of the channel), that we call an ``effective channel'', in which particles with velocities in the direction of a channel can have free paths during a time of at least order $t$, but where there is no real channel. Thus, the convergence to $D t \log(t/\tau)$ is even slower than in the periodic case, because of the variation in width of these effective channels.

\section{Distribution of free path lengths}
\label{results}

We performed numerical simulations of the distribution of free path lengths for a 2D quasiperiodic Lorentz gas obtained using the construction in 
Section~\ref{periodizando}. These were performed for different radii, including the limit case $r=r_c \sim 0.309$, i.e., a Lorentz gas with locally-finite horizon, and  radii $r=0.03$ and $r=0.36$. We used $10^7$  initial conditions, distributed homogeneously in a unit cell in the periodized system and velocities
with unit speed distributed with uniform directions parallel
to the subspace $E$. The orthogonal space $\Eperp$ was taken aligned along the unit vector $(\frac{1}{\phi +2},\frac{\phi}{\phi +2}, \frac{\phi}{\sqrt{\phi +2}})$, where $\phi= \frac{1+\sqrt{5}}{2}$ is the golden ratio.

We have also performed simulations on Sinai Billiard, where we measure the maximum free path length for a fixed direction (slope), as a function of the radius of the obstacle. The simulations were performed with $10^6$ trajectories, with the following slopes: $\frac{1}{\phi \sqrt{\phi+1}}$ (the slope in one of the faces of the unit cell used in the simulations), $\pi$, $\sqrt{2}$, the Liouville constant $\sim 0.110001000000000000000001$ and 10000 random slopes.  

\subsection{Simulation results}

\begin{figure}
\centering

\noindent\begin{tabular}{@{\hspace{0.0em}}c@{\hspace{0.0em}}c@{\hspace{0.0em}}}
   \includegraphics[height=0.6\linewidth, width=120pt]{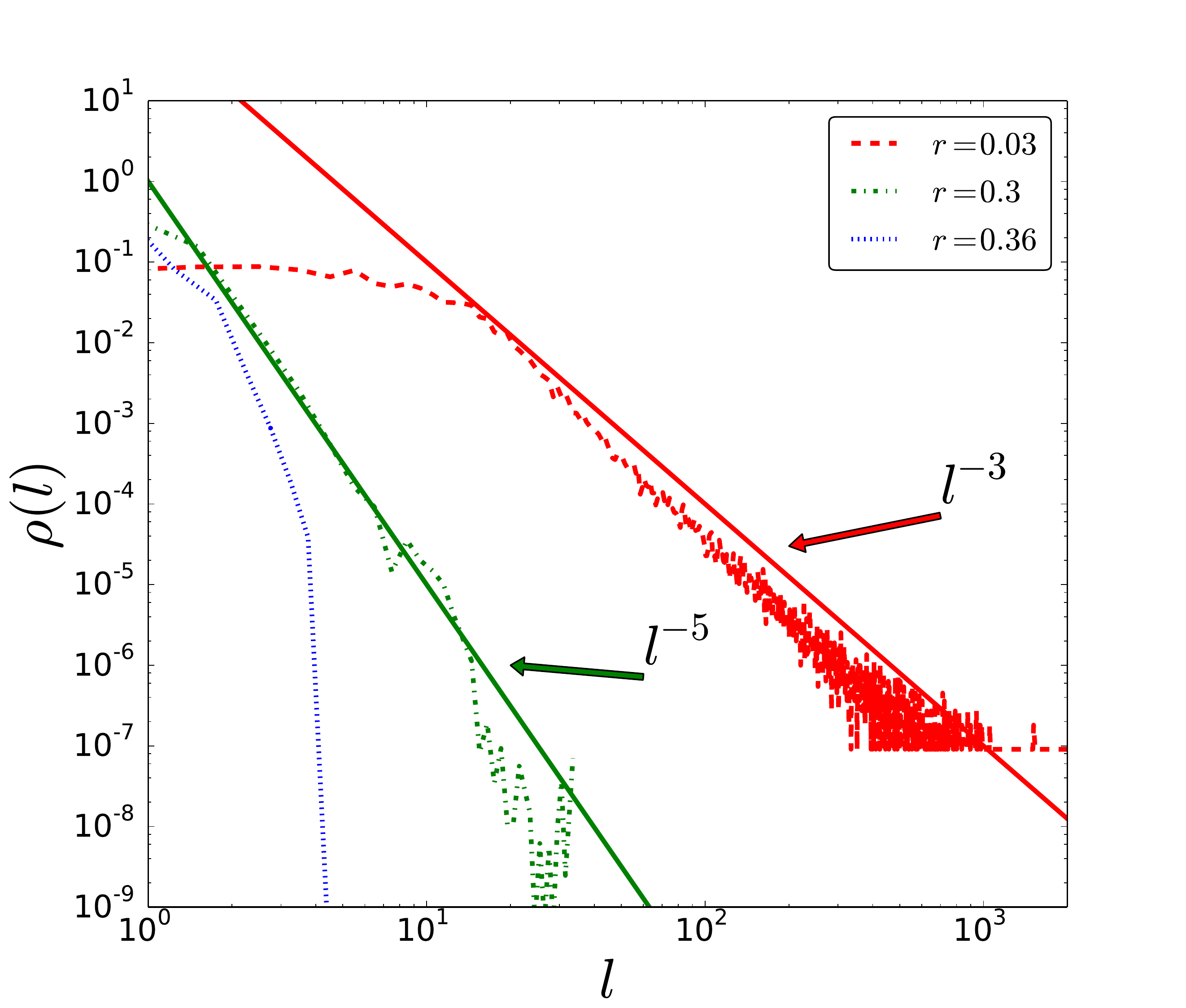} &
   \includegraphics[height=0.6\linewidth, width=120pt]{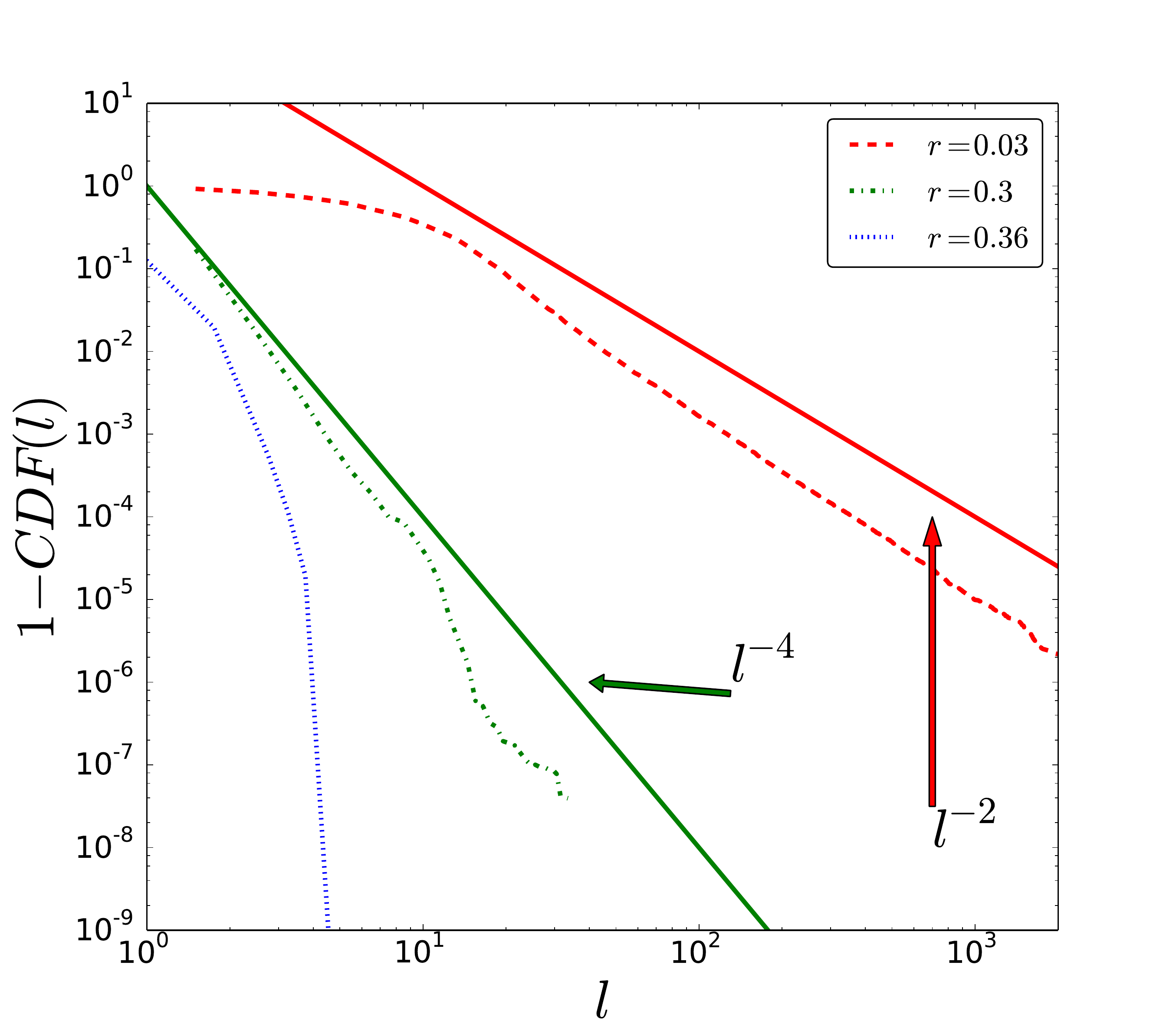} \\
   (a) & (b)
\end{tabular}
 
 \caption{(Color online) (a) Free path length distribution for  radii $r=0.03$ (red (upper) line); $r=0.30$ (green (light gray)); and $r=0.36$ (blue (bottom)). (b) The corresponding cumulative distribution functions with the same color (gray scale) code.}
 \label{fig:simulations}
\end{figure}

The free path distributions obtained from simulations are shown in Fig.~\ref{fig:simulations}. The distribution for $r = r_c$ is well-fitted by a power law with exponent $\alpha=5$, which corresponds to normal diffusion for this case. However, the resolution is not enough good to be conclusive and the calculation is computationally demanding.  Nevertheless, we now give arguments that confirm that the distribution should be a power law with this exponent.   

\subsection{Heuristic argument}

\begin{figure}[ht]
\centering
\includegraphics[width=245pt]{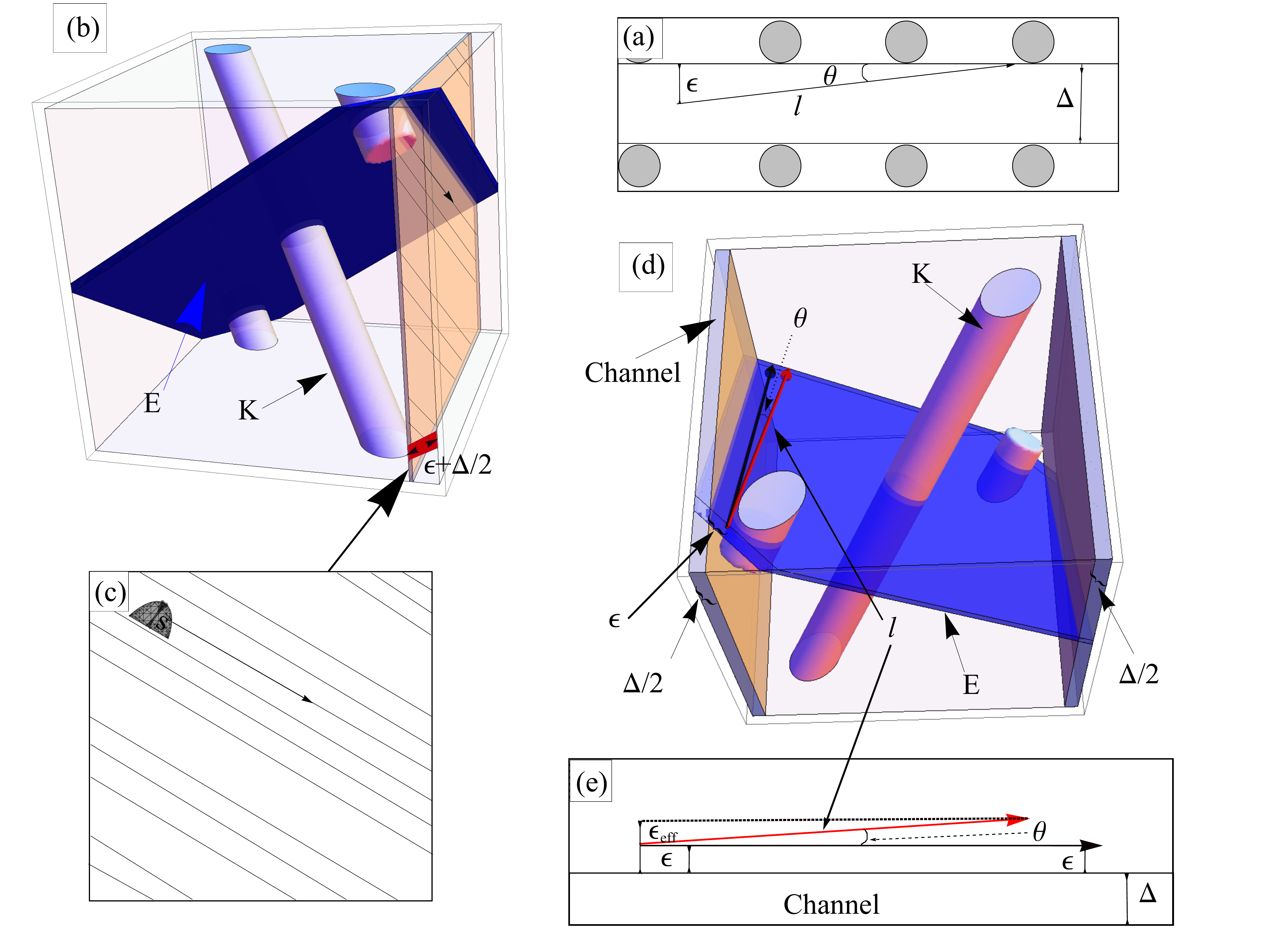}
 \caption{(Color online)(a) Sketch of the geometry to calculate the free path length distribution inside a channel (as in the periodic Lorentz gases). $\epsilon$ measures the distance of the particle from the edge of the channel; 
 $\Delta$ is the width of the channel. (b) and  (c) show how is defined $s$. (d) Another view of the periodized model of figure (b), showing the trajectory of two particles, one with velocity parallel to the direction of the channel (in black), an the other slightly deviated (in red (gray)). This trajectories are also shown in figure (e), where is sketched the geometry to calculate the free path length distribution in the locally finite horizon regime for a 2D quasiperiodic Lorentz gas.}
 \label{fig:channel-distribution}
\end{figure}

Consider a simplification of the periodic case, where a 1D channel corresponds to a system with two parallel lines parallel to the $x$-axis; see Figure~\ref{fig:channel-distribution}(a). In order to calculate the distribution of the length of the first free path, we need to count all the possible initial directions and positions with respect to the channel. If the channel has a width $\Delta$, due to the symmetry of the system, for the position it is enough to consider all  possible initial conditions $(x,y)$ with $x$ an arbitrary constant, and $y \in (0,\Delta/2]$. By symmetry, it is enough to consider angles from $0$ to $\pi/2$ with respect to the $x$-axis; indeed, we are interested in angles close to 0, since our goal is to obtain the distribution for long paths. In this case, as is shown in 
Fig.~\ref{fig:channel-distribution}(a), the length of the free path is $l=\epsilon / \sin(\theta)$, so that $\epsilon= l \cdot \sin(\theta)$. Thus, for angles close to $0$, $\epsilon \sim l \cdot \theta$, the probability $p(l)$ to have a trajectory with length $l$ is
\begin{equation}
p(l) \sim \frac{1}{\theta_m \epsilon_m}\int_{0}^{\theta_m} \int_{0}^{\epsilon_m} \delta (l-\frac{\epsilon}{\theta})d\epsilon d\theta,
\end{equation}
where $\theta_m$ and $\epsilon_m$ are finite constants. Making the change of variables $\xi=\epsilon/\theta$, we obtain 
\begin{equation}
p(l) \sim \frac{1}{\theta_m \epsilon_m}\int_{0}^{\theta_m} \int_{0}^{\frac{\epsilon_m}{\theta}} \delta (l-\xi)\theta d\xi d\theta.
\end{equation}
Using the fact that $\frac{d H(x)}{dx}=\delta(x)$, where $H(x)$ is the Heaviside step function and $\delta(x)$ is the Dirac delta, we obtain
\begin{equation}
\int_{0}^{\theta_m} \int_{0}^{\frac{\epsilon_m}{\theta}} \delta (l-\xi)\theta d\xi d\theta=\int_{0}^{\theta_m} H(\frac{\epsilon_m}{\theta}-l)\theta d\theta.
\end{equation}
Since we assume $\epsilon/l \sim \theta$ and $\epsilon < \epsilon_m$, we have  $\epsilon_m / l > \theta$, giving the approximation
\begin{equation}
p(l) \sim  \int_{0}^{\frac{\epsilon_m} {l}} H(\frac{\epsilon_m}{l}-\theta)\theta d\theta=  \int_{0}^{\frac{\epsilon_m} {l}} \theta d\theta \sim \frac{1}{l^2}.
\end{equation}

In this procedure, we have not considered free paths outside the channel, since in the periodic case the free paths outside a channel are bounded, so the only important contribution to the free path length distribution for long paths is due to the particles inside a channel. However, for the quasiperiodic LG, we have seen that there may be an important contribution from particles outside the channels. 

To analyze this contribution, consider the limiting case, in which the obstacles have the critical radius $r = r_c$. In this case, the only important contribution to the free paths are outside the channels, and changes in position become relevant: for each position, there is a different maximum length of free paths. Placing the particle at a perpendicular distance $\epsilon$ from the channel is equivalent to moving the plane $F$ to the plane $F_\epsilon$. Thus, the intersection of the obstacle with the plane will increase from a point to part of an ellipse; see Figs.~\ref{fig:caricatura2}(2) and  \ref{fig:channel-distribution}(b,c). This ellipse has an effective width of $s=a \epsilon$, with $a$ a constant. Thus, the question becomes: how does the maximum free path length of a particle in a unit square with periodic boundary conditions depend on the radius of the obstacle?

\begin{figure}[ht]
\centering
\includegraphics[width=245pt]{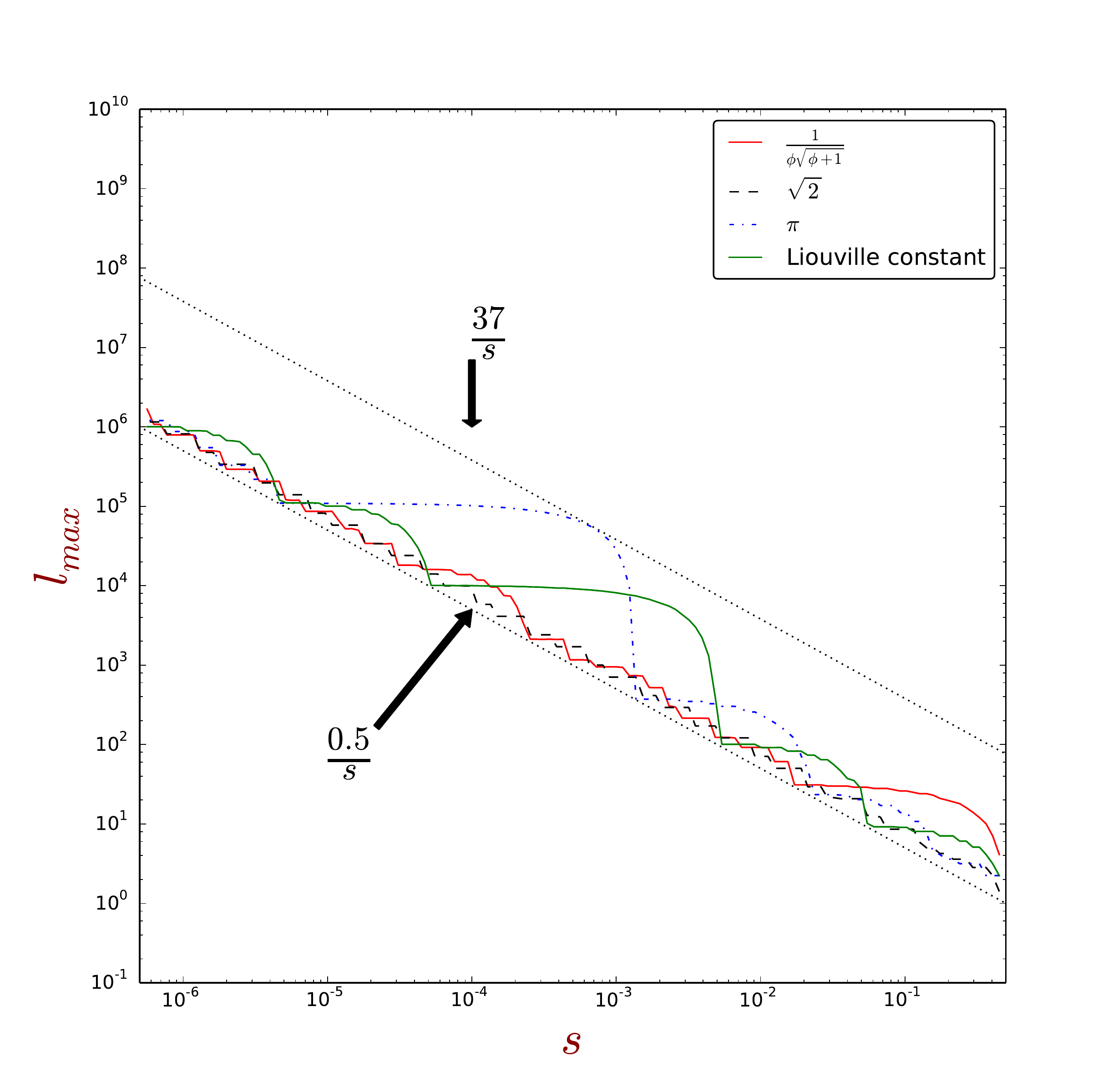}
 \caption{(Color online) Measurement of the maximum free path length as a function of the radius of an obstacle for fixed direction (slope) in a 2D Lorentz gas with the square arrangement. The employed slopes are: $\frac{1}{\phi \sqrt{\phi+1}}$, $\sqrt{2}$, $\pi$ and the Liouville constant $\sim [0.110001000000000000000001]$}
 \label{fig:num-freeflight}
\end{figure}

Figure~\ref{fig:num-freeflight} shows the results of numerical simulations, performed using an efficient algorithm that will be published elsewhere~\cite{Ata-Nicolay-David}, that are well approximated by the function $L(s)= C(s)/s \sim 1/\epsilon$, where $C(s)$ is a bounded function. In order to present these results more clearly, we have included only three slopes. However, we have performed simulations for 10004 different slopes, including 10000 random slopes,  $\frac{1}{\phi \sqrt{\phi+1}}$, $\pi$, $\sqrt{2}$, and the Liouville constant. In any case, $C(s)$ seems to bounded below by the constant $1/2$, while the upper bound depends on the slope; the largest that we have found is for slope $\pi$, and it is around $37$. Thus, we conjecture that $L(s)\sim C(s)/s$, where $C(s)$ is a function bounded below by $1/2$ and bounded above with an upper bound depending on the slope. 

On the other hand, assuming small angles, close to the channel and using trigonometry we obtain  that $l \cos(\theta) \sim l (1-\theta ^2) \sim a/(\epsilon+l\sin(\theta)) \sim a/(\epsilon+l \theta)$; see fig.~\ref{fig:channel-distribution}(d,e). Taking approximations of $\sin(\theta)$ and $\cos(\theta)$ up to second order, we can approximate $l \sim c/\epsilon_{eff}=c/(\epsilon+l \theta)$, with $c$ a constant.

Using this approximation, and the same procedure as for the periodic case, we obtain
\begin{equation}
p(l) \sim \frac{1}{\theta_m \epsilon_m} \int_{0}^{\theta_m} \int_{0}^{\epsilon_m} \delta(l-\frac{c}{\epsilon+l \theta}) d\theta d\epsilon.
\end{equation}
Now we substitute $\xi=c/(\epsilon+l\theta)$, so that $\epsilon=c/\xi-l\theta$, which gives
\begin{equation}
p(l) \sim \frac{1}{\theta_m \epsilon_m} \int_{0}^{\theta_m} H(l-\frac{c}{\epsilon_m+l \theta})H(\frac{c}{l \theta}-l) \frac{1}{l^2} d \theta.
\end{equation}

Assuming $l \gg 1$, we have $1/l-\epsilon_m<0$ and $\theta>1/l (1/l -\epsilon_m)$ or $l-\frac{c}{\epsilon_m+l\theta}>0$. On the other hand, $c/(l \theta)-l>0$ if and only if $\theta<c/l^2$, so the integral becomes
\begin{equation}
p(l) \sim \int_{0}^{\frac{c}{l^2}}\frac{1}{l^2} d l=\frac{1}{l^4}.
\end{equation}

In both cases (in the channel and outside the channel), we have computed the distribution of the free path lengths for the \emph{first} collisions. However, since the distribution of the angles is different for the second collision, i.e., the free paths between two obstacles (rather than starting from a random initial condition in a cell), we still need to calculate the distribution for this case. To do so, assume a distribution of free paths equal to $\rho(l)$, a continuous function of $l$. Since the system is ergodic (this has been proved for periodic systems~\cite{77}, and we assume that it also holds for quasiperiodic ones), we run a simulation with $n$ particles, and stop it at an arbitrary time $t$. The distribution of angles and positions should be homogeneous. So, the free path length distribution at this point should be the same as the first free path length distribution. We ask for this distribution as a function of $\rho(l)$. This problem is equivalent to have a $\rho(l)$ distribution of segments 
of length $l$ on a line, and to choose a point randomly on this line. The probability density that the point particle has a free path of length $l$ if it always moves in a positive direction is $p(l)=\rho(l) \cdot l/2$, so
\begin{equation}
\rho(l)=2\frac{p(l)}{l}.
\end{equation}

Thus, our calculations give $\rho(l) \sim l^{-3}$ and $\rho \sim l^{-5}$  for the distribution of free path lengths inside the channel and outside the channel, respectively. This implies that the distribution should be dominated by $\rho(l) \sim l^{-3}$ when there is an infinite channel, but if the width of the channel $\Delta$ tends to $0$, then we expect the distribution to be  $\rho(l) \sim l^{-5}$. Both results are in good agreement with our numerical simulations, shown in Fig.~\ref{fig:simulations}.

However, this situation does not always occur, because the window to produce the quasiperiodic arrangement is not always totally irrational, even if the space $E$ is. For example, for the Penrose Lorentz gas, this situation does not happen, since the window in one direction forms a rational angle with the channels. In this case, the behavior is similar to that in periodic Lorentz gases. An example of a real quasicrystal arrangement which exhibits this behavior is presented in Ref.~\cite{koca2014twelvefold}. 

Simulations of this system are too slow to study the free path length distribution, but we have computed the directions of the channels using the method of Ref.~\cite{PRL-Ata-David}, and we find that these channels are totally irrational with respect to the window. 

\section{Conclusions}
\label{conclusions}

We have studied the 
properties of Lorentz gases where obstacles are arranged according to quasicrystalline symmetry,
providing a proof of the generic existence of channels for quasiperiodic Lorentz gases in $n$ dimensions for small enough obstacles. This was conjectured in Ref.~\cite{PRL-Ata-David},  but until now the proof was missing. Furthermore, we have given a method to identify these channels and measure their volumes, which is closely related to the super-diffusion coefficient.
We have also proved the existence of a locally-finite horizon regime for quasiperiodic Lorentz gases. This regime occurs at a critical radius $r_c >0$ of obstacles, when the volume of the channels tends to $0$. With this, we have shown that quasiperiodic arrays of obstacles can exhibit three different regimes, finite, infinite and 
locally finite, thus exhibiting richer behaviour than that found in periodic systems. 

In addition, we have performed numerical simulations and heuristic calculations showing that the free path length distribution in the 
locally-finite regime for a 2D quasiperiodic array is asymptotically a power law with  exponent $-5$, instead of $-3$ as in the infinite horizon regime. This allows us to deduce that  diffusion in the locally-finite regime is normal, in agreement with the numerical results of 
Ref.~\cite{PRL-Ata-David}. 
We remark that a similar situation has been found for the Boltzmann-Grad limit for two overlapping periodic lattices  of obstacles with non-commensurate directions \cite{marklof2014power}. 

These results suggest that surprising behaviors can be found in quasiperiodic systems, where not only the mean square displacement, but also the mean free path length is relevant, such as in the calculation of band gaps in photonic crystals \cite{Photonicband,soukoulis2012photonic} recently studied with Lorentz gas models \cite{rousseau2014ray}, or thermal and electrical conductivity \cite{sondheimer1952mean} of quasicrystals. 

\section{Acknowledgements}
We thank Domokos Szász and Jürgen Horbach for useful discussions about the distributions of free path lengths. ASK and MS received support from the DFG withing the Emmy Noether program (grant Schm 2657/2). DPS received financial support from CONACYT Grant CB-101246 and DGAPA-UNAM PAPIIT Grants IN116212 and IN117214.

\bibliographystyle{apsrev}
\bibliography{quasilorentz-freefd2}

\end{document}